# A new distance between DNA sequences

Viswanath.C.Narayanan
Govt. Engineering College, Thrissur, India
Email: narayanan_viswanath@yahoo.com

**Abstract**

**Motivation:** We propose a new distance metric for DNA sequences, which can be defined on any evolutionary Markov model with infinitesimal generator matrix Q. That is the new metric can be defined under existing models such as Jukes-Cantor model, Kimura-2-parameter model, F84 model, GTR model etc. Since our metric does not depend on the form of the generator matrix Q, it can be defined for very general models including those with varying nucleotide substitution rates among lineages. This makes our metric widely applicable. The simulation experiments carried out shows that the new metric, when defined under classical models such as the JC, F84 and Kimura-2-parameter models, performs better than these existing metrics in recovering phylogenetic trees from sequence data. Our simulation experiments also show that the new metric, under a model that allows varying nucleotide substitution rates among lineages, performs equally well or better than its other forms studied.

## 1.Introduction

Distance metrics plays an important role in phylogenetic reconstruction. We have various distance metrics used in phylogenetic analysis which are based on different DNA substitution models such as: Jukes-Cantor [1], Kimura [2,3], Felsenstein [4,5], Hasegawa, Kishino and Yano [6,7], Tamura and Nei [9], Posada [10] and Tavare [8]. In each of these models the substitution process is a continuous-time Markov chain (see [12]) with states {A,C,G,T}, a 4×1 vector of equilibrium probabilities $\pi$ and a 4×4 rate matrix Q. Among all these models the GTR (generalized time reversible) model by Tavare [8] is the most general model in the sense that the rate matrix Q for these model generalizes the rate matrices for the other models. For a more detailed discussion see Huelsenbeck et al. [11]. There are distance metrics under more complex models, which can treat the case of unequal base combinations across lineages, like the one discussed in Galtier and Gouy [13], the paralinear distance [14] and the LogDet distance [15,16].

We propose a distance metric for DNA sequences, which can be defined on any evolutionary Markov model with generator matrix Q. That is it can be defined under the Jukes-Cantor model, the Kimura model, the GTR model etc. This is because the proposed new distance is the average number of mutations (including the possible hidden ones also) that might have occurred during the evolutionary process. For finding this average number, we first assume that the underlying evolutionary process is a Markov process $\Re$ with state space {A,C,G,T} and infinitesimal generator Q. Now identifying the mutations in the evolutionary process with jumps in the Markov process $\Re$, we can trace out all the mutations. The average number of mutations in the evolutionary process is the average number of jumps in $\Re$. This number is calculated and is taken as the distance between the two DNA sequences. Thus we have a new distance between two DNA sequences which gives a direct measure of the average number of mutations including the hidden ones; and which can be defined under a variety of evolutionary models including those which can handle the cases of varying nucleotide substitution rates among lineages.

For testing the efficiency of the new metric, we defined our metric on three existing models namely the Jukes-Cantor model, the Kimura-2-parameter model, and the F84 model and used each of them in phylogenetic tree reconstruction from simulated as well as real data. The efficiency of the new metric in each case is compared with that of the corresponding existing metric in terms of the distance of the recovered tree from the true tree. In many cases, the simulation experiments showed a more efficient performance of our metric over the corresponding existing ones. We also tested the efficiency of our metric, defining it under a fourth model, which can afford varying nucleotide substitution rates across lineages. Our simulation results show that this fourth definition performs equally well with the other three definitions.

The paper is arranged as follows. In section 2 the definition of the distance function and its explicit expression in some particular cases are given. Section 3 discusses the efficiency of the new metric in recovering phylogenetic trees from simulated as well as real data and section 4 concludes the discussion.

## 2 Materials and Methods

### 2.1 Methods
#### 2.1.1 Defining the distance function

Let **F** be the frequency matrix obtained by comparing sites in two DNA sequences X and Y and is given by,

$$\mathbf{F} = \begin{array}{c} \\ \\ X \end{array} \begin{array}{c} \\ A \\ C \\ G \\ T \end{array} \begin{array}{cccc} & & Y & \\ A & C & G & T \\ F_{11} & F_{12} & F_{13} & F_{14} \\ F_{21} & F_{22} & F_{23} & F_{24} \\ F_{31} & F_{32} & F_{33} & F_{34} \\ F_{41} & F_{42} & F_{43} & F_{44} \end{array}$$

For defining a distance between the DNA sequences X and Y, assume that the evolution process, which a cite in X and Y undergoes, is driven by a continuous time Markov chain $\Psi$ with some 4 x 4 generator matrix Q. Identify 'mutations ', which a cite undergoes during the evolution process, with 'jumps' in the process $\Psi$. Let the state space of the process $\Psi$ be {1, 2, 3, 4} and $E_{ij}(t)$ denote the average number of jumps in $\Psi$ which has reached the state j at time t, starting form state i. Then $E_{ij}(t)$ gives the average number of mutations in the interval [0,t) which a cite undergoes during the evolution process between X and Y. Now we can define the average number of mutations in the whole evolution process between X and Y in the time interval [0,t), which is taken as the distance between the sequences X and Y, to be;

$$d_Q(t) = \frac{\sum_{i,j=1}^{4}\left(\left[\frac{F_{ij}+F_{ji}}{2}\right]E_{ij}(t)\right)}{\sum_{i,j=1}^{4}F_{ij}} \qquad \ldots\ldots\ldots\ldots(1)$$

We see that by definition, $d_Q(t)$ is symmetric; $d_Q(t)=0$ if and only if there is no jump in $\Psi$ that is if and only if X=Y. Also the definition of $d_Q(t)$ as the average number of mutations implies that it satisfies the triangle inequality. Hence we have the following theorem.

**Theorem 2.1.1**
When t is fixed, $d_Q(t)$ is a metric.

### 2.1.2 Computation of $E_{ij}(t)$

Let N(s, s+t) be the number of jumps in the process $\Psi$ during the interval (s, s+t). For $1 \le i, j \le 4$, $n \in Z^+$, define, the probabilities;

$$P_{ij}(t,n) = \Pr(\Psi(s+t) = j, N(s,s+t) = n \mid \Psi(s) = i) \qquad \ldots\ldots(2)$$

Then $E_{ij}(t)$ can be obtained as:

$$E_{ij}(t) = \sum_{n=0}^{\infty} nP_{ij}(t,n) \qquad \ldots\ldots(3)$$

The probabilities in equation **(2)** satisfies the following equations (see [17]):

$$P_{ij}(t,0) = \delta_{ij}\exp(Q_{ii}t) \qquad \ldots\ldots(4)$$

and for $n \ge 0$,

$$P_{ij}(t,n+1) = \sum_{\substack{l=1 \\ l \ne i}}^{4}\int_0^t Q_{il}P_{lj}(t-y,n)\exp(Q_{ii}y)dy. \qquad \ldots\ldots(5)$$

Now, let E(t) be the matrix whose (i,j)$^{th}$ entry is $E_{ij}(t)$, $\tilde{E}(t)$ be the diagonal matrix whose i$^{th}$ diagonal entry is $\exp(Q_{ii}t)$, and J be the matrix whose (i,j)$^{th}$ entry is $Q_{ij}$, for $i \ne j$, and whose diagonal entries are zeros. Then equation **(5)** can be transformed into the matrix integral equation;

$$E(t) = \int_0^t \tilde{E}(y)JE(t-y)dy + \int_0^t \tilde{E}(y)JP(t-y)dy. \qquad \ldots\ldots(6)$$

Solving this equation gives the required $E_{ij}(t)$ s'.

**Example 2.1.1**

If we take the generator matrix Q as

$$Q = \begin{bmatrix} -\alpha-\beta-\gamma & \alpha & \beta & \gamma \\ \alpha & -\alpha-\beta-\gamma & \gamma & \beta \\ \beta & \gamma & -\alpha-\beta-\gamma & \alpha \\ \gamma & \beta & \alpha & -\alpha-\beta-\gamma \end{bmatrix},$$

which is the rate matrix in the Kimura 3-parameter model, then $E_{ij}(t) = E_{ji}(t)$ for all i,j and

$$E_{ii}(t) = \exp(-(\alpha+\beta+\gamma)t)\sum_{n=2}^{\infty} H_n \frac{t^n}{(n-1)!},$$

$$E_{12}(t) = E_{34}(t) = \exp(-(\alpha+\beta+\gamma)t)\sum_{n=1}^{\infty} \overline{H}_n \frac{t^n}{(n-1)!},$$

$$E_{13}(t) = E_{24}(t) = \exp(-(\alpha+\beta+\gamma)t)\sum_{n=1}^{\infty} \tilde{H}_n \frac{t^n}{(n-1)!},$$

$$E_{14}(t) = E_{23}(t) = \exp(-(\alpha+\beta+\gamma)t)\sum_{n=1}^{\infty} \hat{H}_n \frac{t^n}{(n-1)!},$$

with

$\overline{H}_1 = \alpha, \tilde{H}_1 = \beta, \hat{H}_1 = \gamma,$

$H_2 = 2\gamma\beta, \tilde{H}_2 = 2\alpha\gamma, \hat{H}_2 = 2\alpha\beta,$

$H_{n+1} = \alpha\overline{H}_n + \beta\tilde{H}_n + \gamma\hat{H}_n \; ; \; n \geq 1,$

$\overline{H}_{n+1} = \alpha H_n + \gamma\tilde{H}_n + \beta\hat{H}_n \; ; \; n \geq 2,$

$\tilde{H}_{n+1} = \beta H_n + \gamma\overline{H}_n + \alpha\hat{H}_n \; ; \; n \geq 2,$

$\hat{H}_{n+1} = \gamma H_n + \beta\overline{H}_n + \alpha\tilde{H}_n \; ; \; n \geq 2.$

For the Jukes-Cantor model, which is the particular case of the above model with $\alpha = \beta = \gamma = \frac{\mu}{4}$, we have more explicit expressions of $E_{ij}(t)$'s as:

$$E_{ij}(t) = \left(\frac{t\mu}{4}\right)\left(\exp\left(\frac{-3t\mu}{4}\right)\right)\left\{1 + 2\frac{\left(\frac{t\mu}{4}\right)}{1!} + 7\frac{\left(\frac{t\mu}{4}\right)^2}{2!} + 20\frac{\left(\frac{t\mu}{4}\right)^3}{3!} + 61\frac{\left(\frac{t\mu}{4}\right)^4}{4!} + \cdots\right\},$$

for $1 \leq i, j \leq 4, i \neq j$, and for $1 \leq i \leq 4$,

$$E_{ii}(t) = \left(\frac{t\mu}{4}\right)\left(\exp\left(\frac{-3t\mu}{4}\right)\right)\left\{3\frac{\left(\frac{t\mu}{4}\right)}{1!} + 6\frac{\left(\frac{t\mu}{4}\right)^2}{2!} + 21\frac{\left(\frac{t\mu}{4}\right)^3}{3!} + 60\frac{\left(\frac{t\mu}{4}\right)^4}{4!} + \cdots\right\}.$$

### 2.2 Materials.

Software packages Mesquite [19] and Seq-Gen.v1.3.2 [20] are used for simulating sequence data on given random and real trees respectively. Real trees are taken from TreeFam Database [21,22]. MBEToolbox 2.0 [23] is used for getting the distance matrices using the new metric. For getting distance matrices using the JC, K2P, F84 distances, and for phylogenetic tree construction with the new as well as the existing metrics, PHYLIP 3.66 [24] is used. For comparing the phylogenetic trees recovered from the sequence data with the original tree, we used the package TOPD/FMTS [25].

## 3. Phylogenetic tree reconstruction based on $d_Q(t)$.

To substantiate the claim that the new metric improves the existing metrics such as the Jukes-Cantor metric, F84 metric and Kimura-2-parameter metric, we conducted simulation experiments with simulated as well as real data. The details of these experiments are as follows.

### 3.1 Selection of the evolutionary model

Clearly, the challenge with the metric $d_Q(t)$ is the selection of the generator matrix Q. We defined $d_Q(t)$ in four different ways. The first three definitions were based on the Jukes-Cantor, F84 and Kimura-2-parameter models respectively. The fourth one is obtained first by defining the matrices $P$, $\tilde{P}$ and G as:

$$P_{ij} = \frac{F_{ij}}{\sum_{k=1}^{4} F_{ik}} \quad, \tilde{P}_{ij} = \frac{F_{ji}}{\sum_{k=1}^{4} F_{ki}}, \quad G = \frac{1}{2}(P + \tilde{P});$$

and then taking $Q_{ij} = G_{ij}$, for $i \neq j$. For easy identification of each of these definitions, let us rename $d_Q(t)$ in each of these cases as $d_{QJ}(t)$, $d_{QF}(t)$, $d_{QK}(t)$ and $d_{QD}(t)$ respectively.

### 3.2 Simulation results

We conducted three types of experiments. The basic nature of each type was to select trees first, either real or random, and then try to recover the phylogenies from DNA sequence data, either simulated or actual, using neighbor-joining method together with the new as well as existing metrics. The recovered trees are then compared with the original tree, in terms of the nodal and split distances, using the software TOPD/FMTS.

In the type1 experiment, using the software package Mesquite, we selected 100 random trees with number of species ranging from 25 to 124 and then with the composite simulation model, simulated DNA sequences each of length 1000 bases. We then tried to recover the original tree from the simulated sequence data applying the neighbor-joining method with the new as well as the existing metrics. Comparing the recovered trees with the original tree, the following results were obtained. The average nodal distances using the new metrics $d_{QJ}(t)$, $d_{QF}(t)$ and $d_{QK}(t)$ were 2.5776, 2.6919, 2.6991 respectively; whereas the average nodal distances using the corresponding existing metrics were 4.7693, 10.1901, 9.752 respectively. These average values shows that the new metric can improve the existing metrics and also that the new metric performs slightly better under the Jukes-Cantor model. Figures 1-3 shows the comparison of the new metric with the corresponding existing metric in terms of nodal distance. Figure 4 shows the comparison of the metrics $d_{QJ}(t)$, $d_{QF}(t)$ and $d_{QK}(t)$, which are various forms of the new metric under different models, in terms of the nodal distance.. The average split distance with all the three existing metrics was found to be 1, the maximum possible value. The average split distances using the new metrics $d_{QJ}(t)$, $d_{QF}(t)$ and $d_{QK}(t)$ was 0.3735, 0.4516, 0.4478 respectively, pointing to the better performance of the new metric. Figure 5 shows the comparison of the various forms of the new metric under different models in terms of the split distance. For the calculation of the new metric, here the time t is taken as 0.5.

For the second type of experiments, we selected 25 real trees, with number of sequences varying between 4 and 63, from TreeFam Database. DNA sequences are then simulated for these trees using Seq-Gen. For each tree, we simulated three types of DNA sequences by multiplying the branch lengths by 1,10 and 100 respectively. In the first case, that is when we used the sequences simulated on trees with same branch lengths as obtained from TreeFam, we found our metric performing less efficiently as compared to the existing ones. Here we observed that the simulated sequences were too close to each other; so that the nondiagonal elements in each raw of the frequency matrix F were weaker compared to its diagonal elements. Concluding that this may be the reason for the poor performance of our metric, we then simulated sequences by multiplying the branch lengths of each tree with 10. Since this increased the difference between the simulated sequences and strengthened the nondiagonal elements of F, this time we anticipated a better performance of our metric over the existing metrics. Except for the Kimura-2-parameter distance, we got the expected result. That is the new metric, though narrowly, improved the Jukes-Cantor and F84 metrics. A third simulation is then carried out by multiplying the branch lengths with 100. This time also the new metric couldn't improve the Kimura-2-parameter metric; but the improvement brought by the new metric to the Jukes-Cantor and F84 metrics became more evident. These experimental results in terms of average nodal and split distances are given in table1. Figures 6-8 shows the comparison results between the new and Jukes-Cantor metrics in terms of the nodal distance, when branch lengths are multiplied by 1,10 and 100 respectively. Figures 9-11 show the split distance comparison results between the new and Jukes-Cantor metrics, as branch lengths are multiplied by 1,10 and 100 respectively. Figures 12-17 show the above results for the new and Felsenstein 84 metrics and these results for the new and Kimura-2-parameter distance metrics are given in figures 18-23. For the calculation of the new metric, here the time t is taken as 1.5.

The third experiment was based on a known tree of 11 vertebrate species; the same tree studied in Russo et al. [18]. The 11 species and their GenBank Accession numbers are given in Table 2. We tried to recover this known tree from nucleotide sequence data using the existing as well as the new metrics. Distance comparison is done as in the above two experiments. This experiment also suggested that the new metric improves the existing metrics. The results are given in Table 3. According to this table, the efficiency of the new metric in recovering correct phylogenies from real sequence data is comparatively maximum, when it is defined with the Kimura-2-parameter model. For the calculation of the new metric, here the time t is taken as 1.5.

All the three type of experiments show that the new metric, when defined on the existing evolutionary models such as the Jukes-Cantor, Felsenstein 84, and Kimura-2-parameter models, recovers better phylogenetic trees from the sequence data than the existing metrics. The results also show that the new metric, defined on the fourth model, which can afford varying nucleotide substitution rates across lineages, performs equally well to the best performer among the other three new metrics.

**Table 1:** Average distance comparison of the new metric with the existing metrics

|  |  | Original branch length | Branch length x 10 | Branch length x 100 |
|---|---|---|---|---|
| Average nodal distance | J-C metric | 1.259 | 1.7648 | 4.0657 |
|  | $d_{QJ}(t)$ | 1.3762 | 1.7172 | 2.4302 |
|  | F84 metric | 1.0843 | 3.0020 | 4.6028 |
|  | $d_{QF}(t)$ | 1.6651 | 2.2334 | 3.0511 |
|  | K2P metric | 1.0812 | 2.7714 | 4.3869 |
|  | $d_{QK}(t)$ | 5.0673 | 5.4532 | 7.1264 |
|  | $d_{QD}(t)$ | 1.4781 | 1.7747 | 2.4256 |
| Average split distance | J-C metric | 0.1646 | 0.3132 | 0.8639 |
|  | $d_{QJ}(t)$ | 0.1962 | 0.2903 | 0.5362 |
|  | F84 metric | 0.1911 | 0.6924 | 0.9247 |
|  | $d_{QF}(t)$ | 0.3351 | 0.4401 | 0.6764 |
|  | K2P metric | 0.1518 | 0.6330 | 0.9267 |
|  | $d_{QK}(t)$ | 0.7262 | 0.8222 | 0.9229 |
|  | $d_{QD}(t)$ | 0.2084 | 0.3004 | 0.5346 |

**Table 2:** The 11 vertebrate species and their GenBank accession numbers.

| Species | GenBank accession numbers |
|---|---|
| Balaenoptera physalus | X61145 |
| Balaenoptera musculus | X72204 |
| Bos taurus | V00654 |
| Mus musculus | V00711 |
| Rattus norvegicus | X14848 |
| Didelphis virginiana | Z29573 |
| Gallus gallus | X52392 |
| Xenopus laevis | M10217 |
| Oncorhynchus mykiss | L29771 |
| Crossostoma lacustre | M91245 |
| Cyprinus carpio | X61010 |

**Table 3:** Distance comparison of the new metric with existing metrics based on a known tree of 11 vertebrate species

|              | Nodal distance | Split distance |
|--------------|----------------|----------------|
| J-C metric   | 2.5154         | 0.75           |
| $d_{QJ}(t)$  | 1.9909         | 0.75           |
| F84 metric   | 2.7634         | 1              |
| $d_{QF}(t)$  | 2.0538         | 0.75           |
| K2P metric   | 3.1909         | 1              |
| $d_{QK}(t)$  | 0.7135         | 0.125          |
| $d_{QD}(t)$  | 1.5954         | 0.625          |

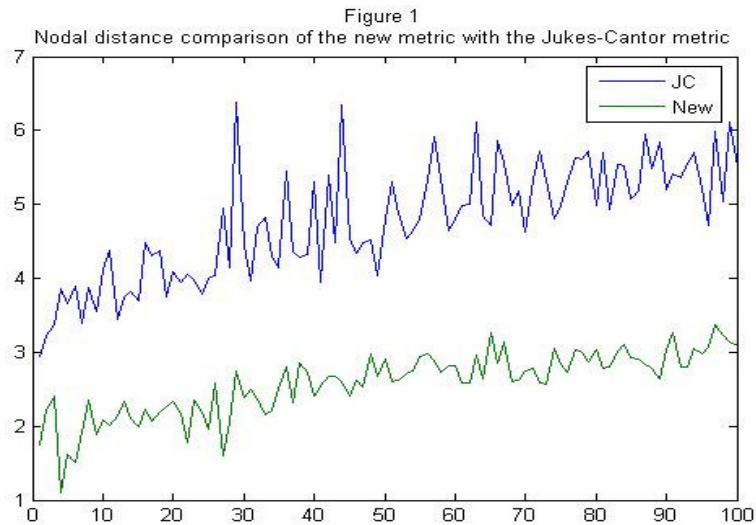

Figure 1
Nodal distance comparison of the new metric with the Jukes-Cantor metric

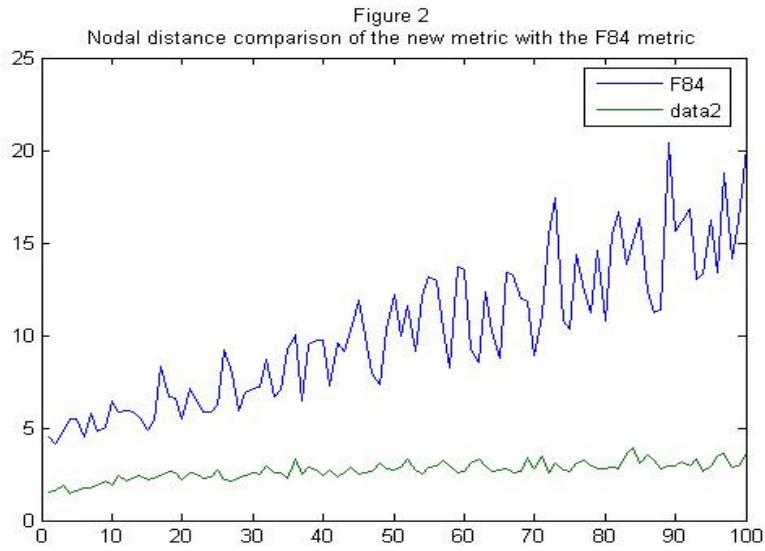

Figure 2
Nodal distance comparison of the new metric with the F84 metric

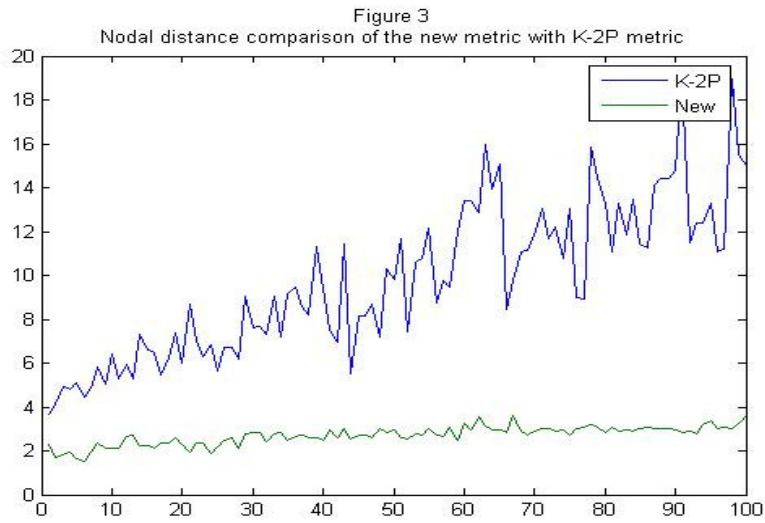

Figure 3
Nodal distance comparison of the new metric with K-2P metric

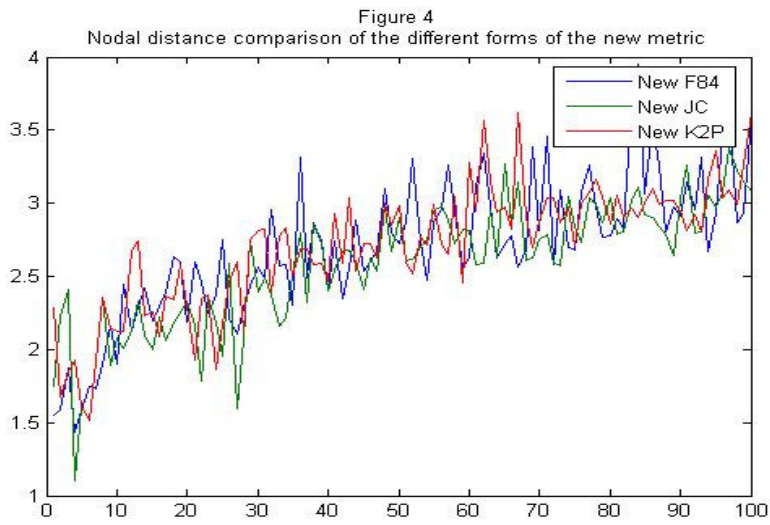

Figure 4
Nodal distance comparison of the different forms of the new metric

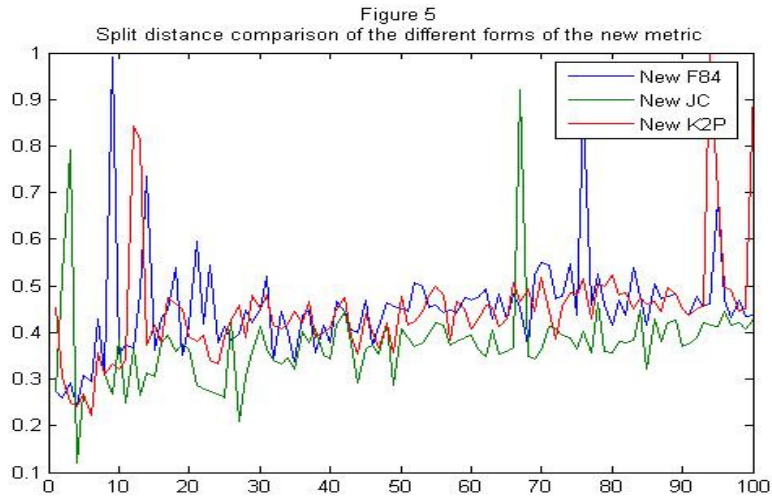

Figure6

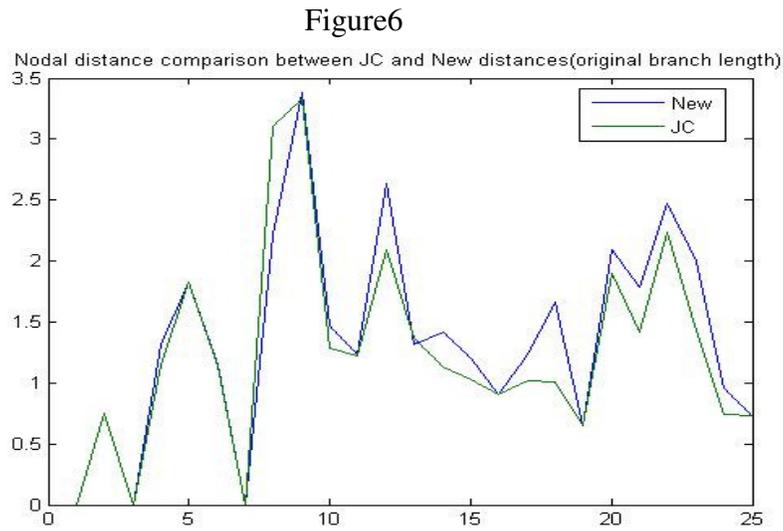

Figure7

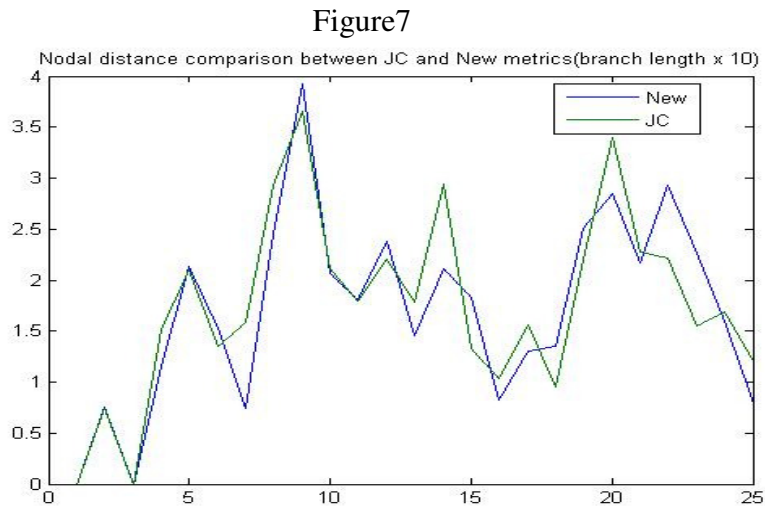

Figure8

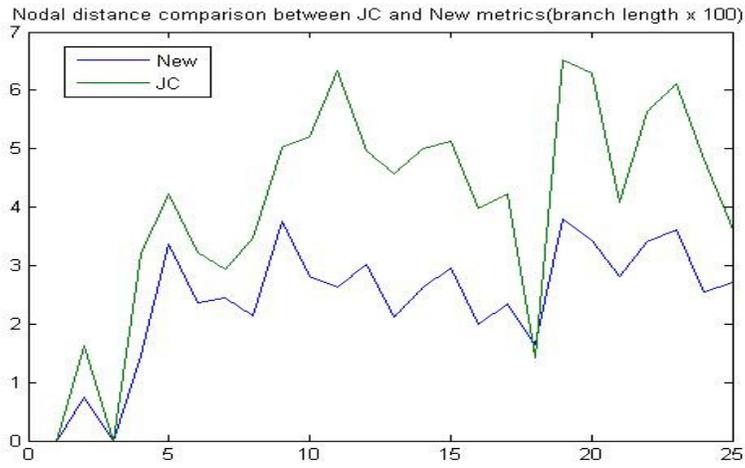

Figure9

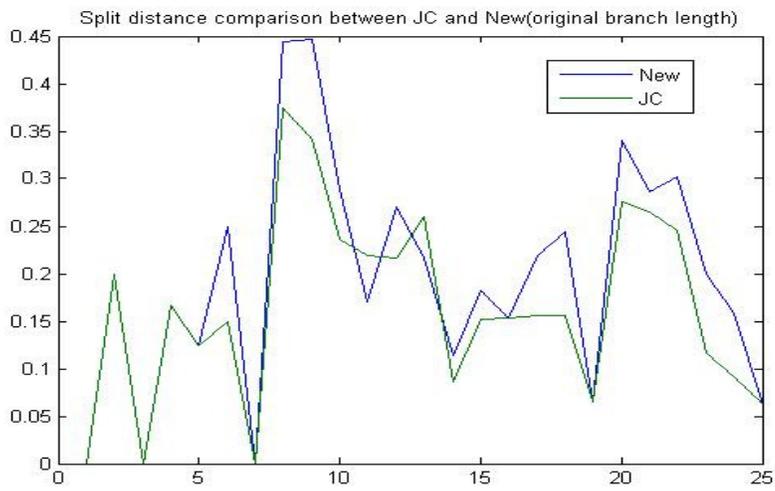

Figure10

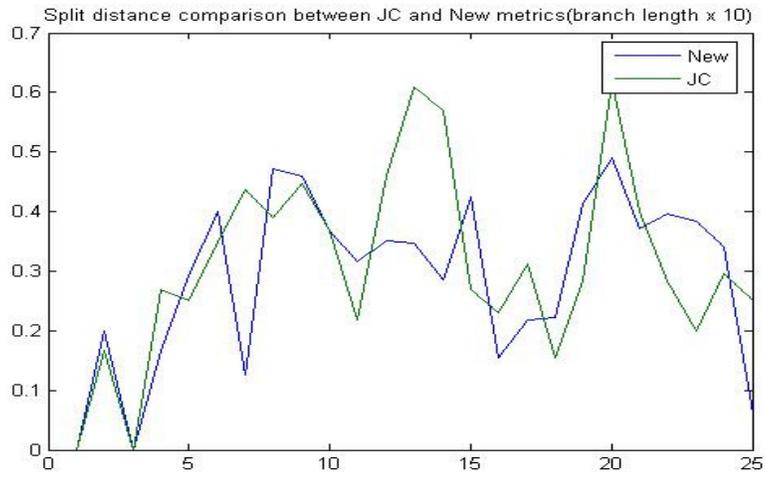

Figure11

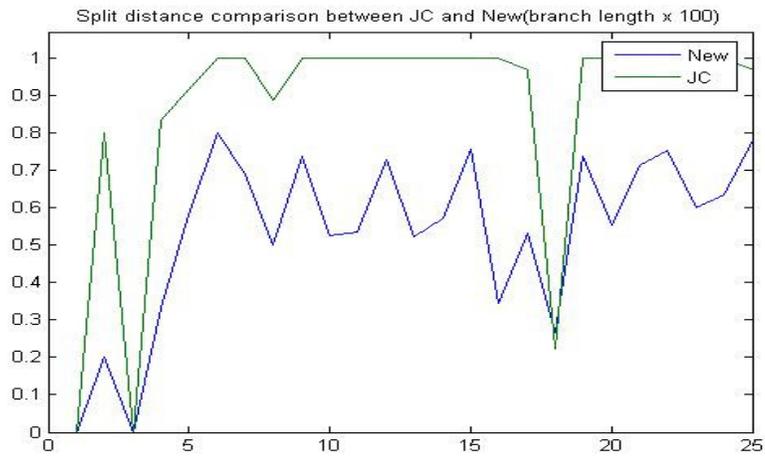

Figure12

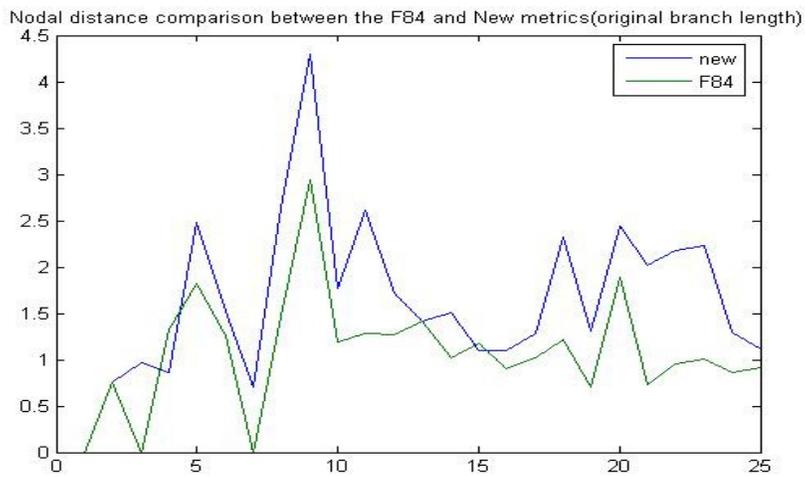

Figure13

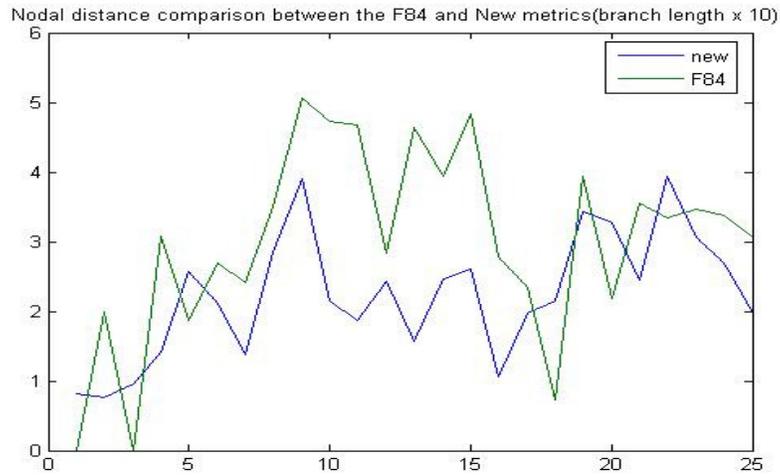

Figure14

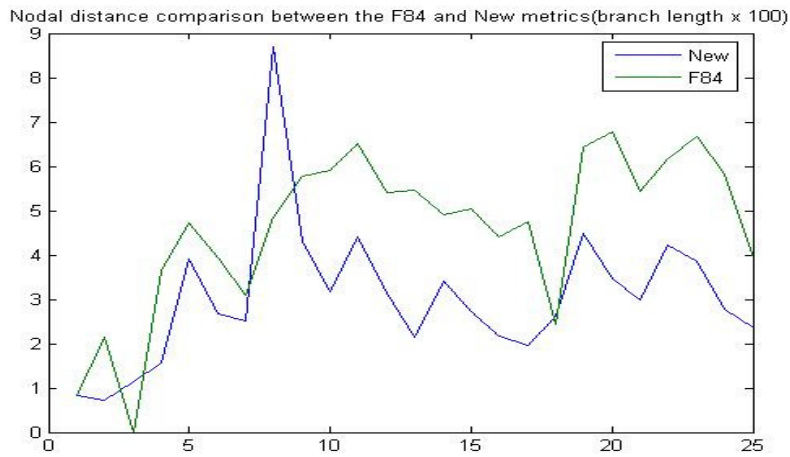

Figure15

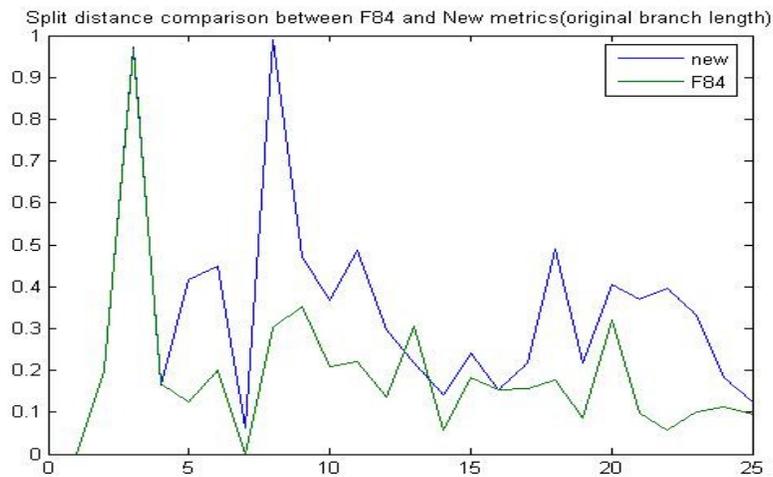

Figure16

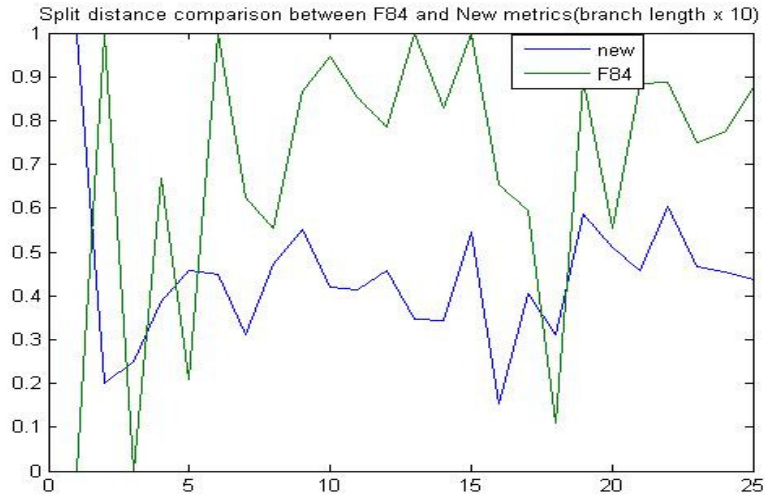

Figure17

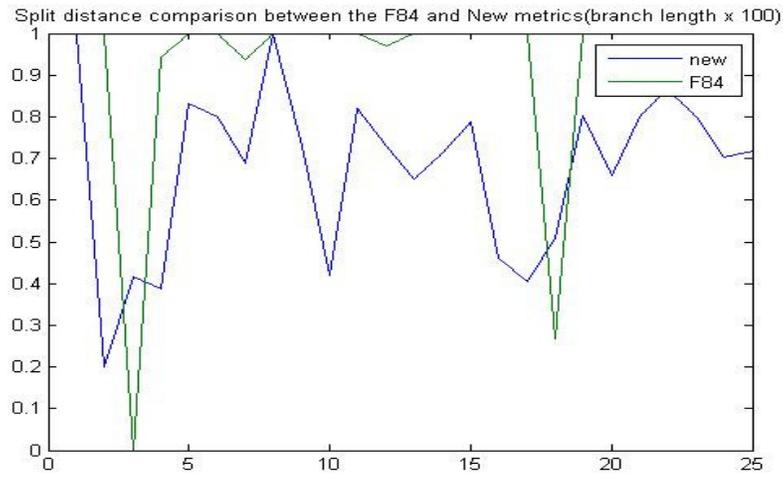

Figure18

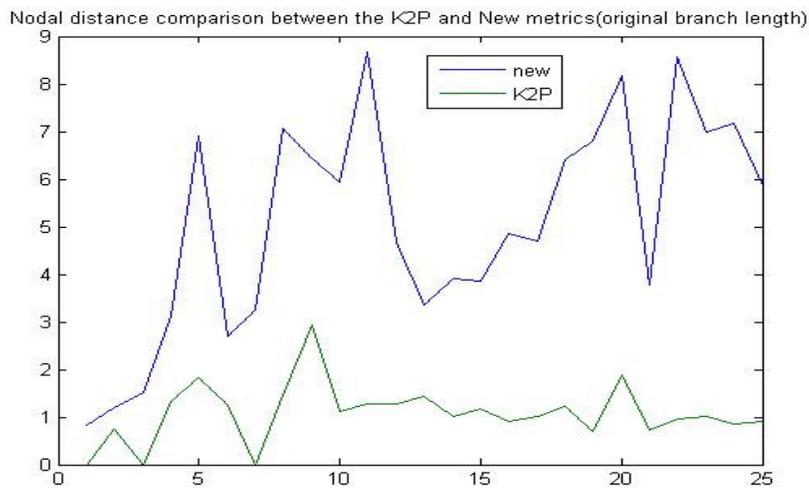

Figure19

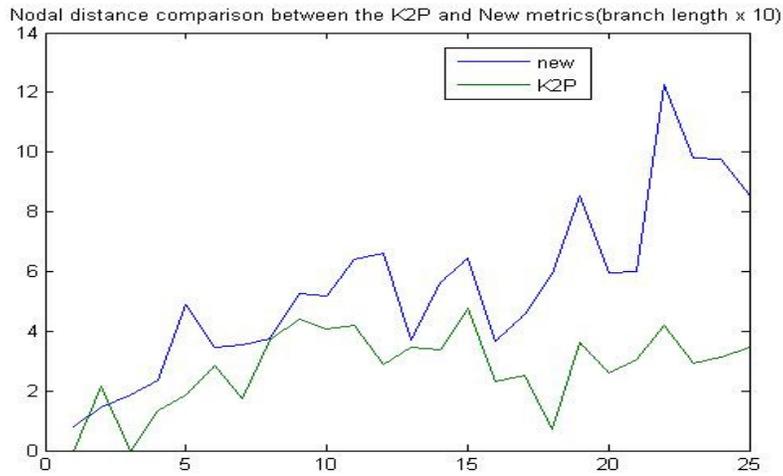

Figure20

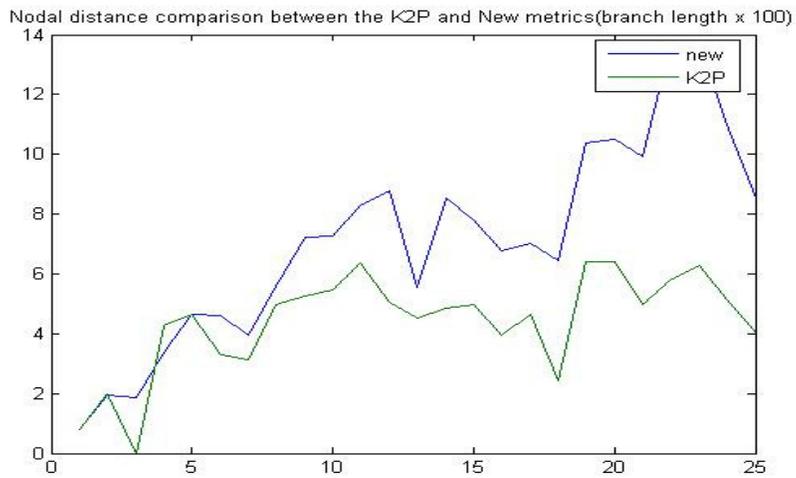

Figure21

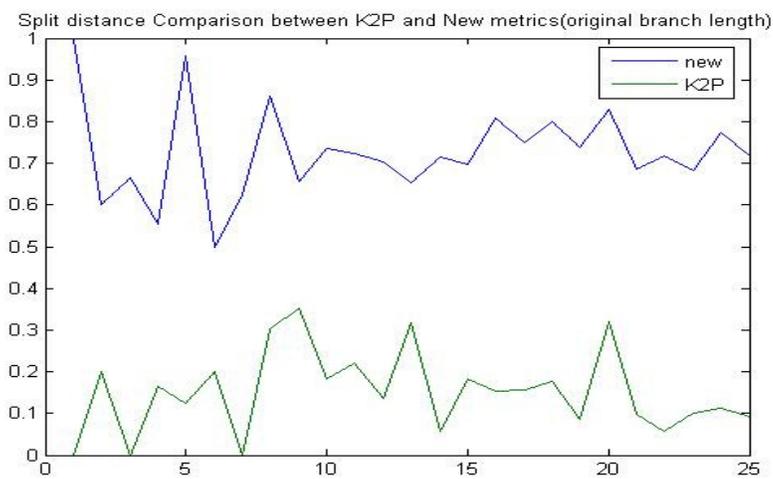

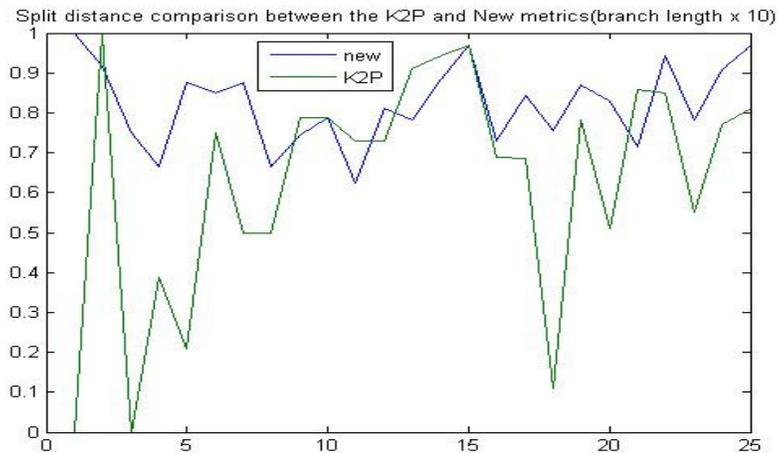

Figure22

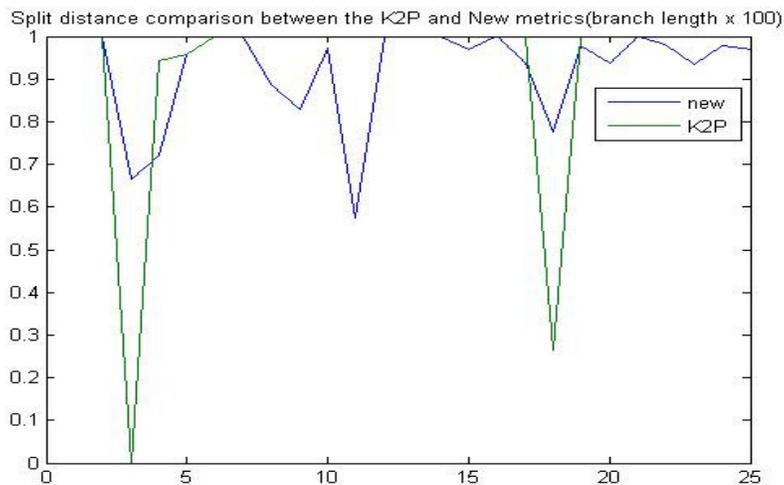

Figure23

## 4. Conclusion

The distance metric $d_Q(t)$ defined here has the following advantages.

1. It can be defined under any evolutionary Markov model with rate matrix Q; without any further assumption on the matrix Q. This makes it definable under very general models and therefore is widely applicable.

2. It measures the average number of base substitutions including mutations.

3. When defined under classical models such as the JC, F84 and Kimura-2-parameter models, it recovered good phylogenetic trees from simulated as well as real sequence data on a given tree.

4. The simulation experiments shows that the new metric, when defined under the JC, K2P and F84 models, improves these existing metrics as far as recovering phylogenetic trees from sequence data is concerned.

5. In the simulation experiments carried out, the definition of the new metric, under a model that allows varying nucleotide substitution rates among lineages, showed a second best performance among the various definitions of the new metric.